\documentclass[useAMS, usenatbib, usegraphicx]{mn2e}
\usepackage{times}

\title[The CGPSE 1420 and 408 MHz Catalogue]{A sharper view of the
  outer Galaxy at 1420 and 408 MHz from the Canadian Galactic Plane
  Survey II: The catalogue of extended radio sources}

\author[C. R. Kerton, J. Murphy \&
  J. Patterson]{C. R. Kerton\thanks{E-mail: kerton@iastate.edu},
  J. Murphy and J. Patterson\\ Iowa State University, Department of
  Physics and Astronomy, Ames, IA, 50011, USA}

\begin{document}

\maketitle

\begin{abstract}
A new catalogue of extended radio sources has been prepared based on
arcminute-resolution 1420 MHz images from the Canadian Galactic Plane
Survey (CGPS). The new catalogue provides both 1420 MHz and 408 MHz flux
density measurements on sources found near
the Galactic plane in the second quadrant of our Galaxy. In addition
cross-identifications are made with other major radio catalogues and
information is provided to facilitate the recovery of CGPS image data
associated with each catalogued source. Numerous new radio sources are
identified and the catalogue provides a comprehensive summary of both
newly discovered and previously known \mbox{H\,{\sc ii}} regions and
supernova remnants in the outer Galaxy. The catalogue should be of use both for
synoptic studies of Galactic structure and for placing higher
resolution observations, at radio and other wavelengths, in context. 
\end{abstract}

\begin{keywords}
surveys -- catalogues -- Galaxy: disc -- radio continuum: general.
\end{keywords}

\section{Introduction} \label{sec:intro}

The Canadian Galactic Plane Survey (CGPS; \citealt{tay03}) has recently
completed observations of the second quadrant of the Galaxy at 1420
and 408 MHz. The images from this survey provide an unprecedented view
of the extended radio continuum emission in this portion of the Galaxy
at arcminute-scale resolution. This paper presents positional and flux
density information on all extended sources identified in the CGPS
data covering $90\degr < l < 175\degr$ between $-3\fdg6 < b <
+5\fdg6$. Cross-identifications with other major catalogues of
extended radio sources and optical \mbox{H\,{\sc ii}} regions are
provided along with mosaic codes that will facilitate the acquisition
of CGPS data for individual sources from the Canadian Astronomy Data
Centre (CADC). The catalogue is especially well suited as the starting
point for large-scale studies of the outer Galaxy, especially those
focusing on the structure of the Perseus and Outer spiral arms, and for
placing small-scale observations in their proper context.

\begin{table*}
\centering
\begin{minipage}{180mm}
\caption{CGPS extended sources -- position, angular size and flux density}
\label{tbl:cat1}
\begin{tabular}{lccccccccccc}

\hline
CGPSE & $l$ & $b$ & RA (J2000) & DEC (J2000) & $\Theta_\mathrm{maj}$
& $\Theta_\mathrm{min}$ & F$_\nu$ (1420) & $\sigma$ (1420) & F$_\nu$
(408)    & $\sigma$ (408) & Mosaic Code \\
      & $\degr$ & $\degr$  & h m s  & $\degr$ \/ $\arcmin$ & arcmin &
arcmin  &   (mJy)        &       (mJy)     &    (mJy)         &
(mJy) & \\
\hline
  1 &  90.240 & $1.720$  & 21 05 18  &  49 40 &    5.70  &   5.70  & $7.10\times10^2$  &  $2.0\times10^1$  &  $5.39\times10^2$  & $2.6\times10^1$  & ML2 \\
  2 &  90.310 & $1.490$  & 21 06 38  &  49 34 &    5.16  &   4.56  & $1.08\times10^2$  &  $4.7\times10^0$  &  $5.00\times10^1$  & $8.5\times10^0$  &   ML2 \\
  3 &  90.495 & $-2.915$ & 21 26 21  &  46 38 &    6.72  &   1.50  & $1.10\times10^2$  &  $4.1\times10^0$  &  $2.00\times10^2$  & $5.5\times10^0$  &  MK1 \\
  4 &  91.020 & $1.900$  & 21 07 44  &  50 22 &   31.44  &  11.94  & $4.10\times10^3$  &  $3.4\times10^1$  &  $2.70\times10^3$  & $1.4\times10^2$  &  MK2 \\
  5 &  91.115 & $1.585$  & 21 09 35  &  50 14 &    5.16  &   5.10  & $1.89\times10^3$  &  $1.7\times10^1$  &  $2.05\times10^3$  & $3.5\times10^1$  &   MK2 \\
  6 &  90.970 & $1.550$  & 21 09 08  &  50 06 &   11.70  &   8.58  & $4.53\times10^2$  &  $1.6\times10^1$  &  $2.75\times10^2$  & $7.5\times10^1$  &  MK2 \\
  7 &  91.550 & $0.975$  & 21 14 13  &  50 07 &    6.78  &   5.22  & $2.38\times10^2$  &  $9.5\times10^0$  &  $7.92\times10^1$  & $1.1\times10^0$  &  MK2 \\
  8 &  91.790 & $-0.085$ & 21 19 58  &  49 33 &    7.08  &   5.88  & $8.02\times10^1$  &  $8.5\times10^0$  &          $\cdots$  &        $\cdots$  &  MK1 \\
  9 &  92.155 & $3.080$  & 21 07 03  &  52 00 &   73.26  &  20.52  & $1.59\times10^3$  &  $1.4\times10^2$  &  $1.87\times10^3$  & $1.7\times10^2$  &  MK2 \\
 10 &  92.240 & $1.560$  & 21 14 34  &  51 02 &    7.08  &   1.68  & $9.94\times10^1$  &  $6.9\times10^0$  &  $1.68\times10^2$  & $3.1\times10^1$  &  MK2 \\
\hline
\end{tabular}
\medskip
Table~\ref{tbl:cat1} is presented in its entirety in the electronic
edition of the journal. 
\end{minipage}
\end{table*}

Prior to the CGPS the best synoptic view of the radio continuum
emission at centimetre wavelengths in the outer Galaxy was provided by
a series of surveys done by the Effelsberg 100-m telescope at 9-arcmin
resolution. The surveys were summarized in the \citet{kal80} and
\citet*{rrf97} catalogues (KR and RRF respectively). Another very
useful summary is the catalogue of outer Galaxy radio sources within
$93\degr < l < 163\degr$,  $|b| < 4\degr$ compiled by \citet{fic86}, as
part of his VLA ``snapshot'' survey of point sources in the KR
catalogue. The \citet{fic86} catalogue includes all of the KR
sources along with extended objects observed by the Effelsberg 100-m
telescope but not noted in the KR catalogue.

\begin{table*}
\centering
\begin{minipage}{160mm}
\caption{CGPS extended sources -- cross-identifications and notes}
\label{tbl:cat2}
\begin{tabular}{lccccccl}
\hline
CGPSE & KR & Kothes et al. (2006) & Sh-2  & RRF & F3R & Paladini et al. (2003) & Notes \\
\hline
1   & $\cdots$ & $\cdots$   & 121      & 825      & 2906     &    781   & $\cdots$    \\
2   & $\cdots$ & $\cdots$   & $\cdots$ & $\cdots$ & 2910     & $\cdots$ & $\cdots$    \\
3   & $\cdots$ & $\cdots$   & $\cdots$ & $\cdots$ & $\cdots$ & $\cdots$ & B3~2124+464 \\
4   & $\cdots$ & $\cdots$   & $\cdots$ & $\cdots$ & $\cdots$ & $\cdots$ & BG~2107+49, north plume  \\
5   & $\cdots$ & $\cdots$   & $\cdots$ & 835      & 2930     & $\cdots$ & BG~2107+49, compact core \\
6   & $\cdots$ & $\cdots$   & $\cdots$ & $\cdots$ & 2927     & $\cdots$ & BG~2107+49, south plume  \\
7   & $\cdots$ & $\cdots$   & $\cdots$ & $\cdots$ & 2941     & $\cdots$ &  $\cdots$   \\
8   & $\cdots$ & $\cdots$   & $\cdots$ & $\cdots$ & 2945     & $\cdots$ &  $\cdots$   \\
9   & 1        & $\cdots$   & $\cdots$ & $\cdots$ & $\cdots$ & $\cdots$ &  filaments probably associated with CGPSE~14 (KR~1) \\
10  & $\cdots$ & $\cdots$   & $\cdots$ & 850      & 2957     & $\cdots$ &  $\cdots$ \\
\hline
\end{tabular}
\medskip
Table~\ref{tbl:cat2} is presented in its entirety in the electronic
edition of the journal.
\end{minipage}
\end{table*}

Paper 1 in this series \citep{ker06} used CGPS data to look at sources listed 
in the KR catalogue focusing on those sources misclassified as
extended sources, and thus not observed by \citet{fic86}, along with
point sources with flat or rising spectra between 408 and 1420
MHz. This paper presents a catalogue of {\it all} extended emission
features seen in the CGPS radio continuum data and thus provides an
expanded and updated version of the \citet{fic86}
catalogue. Specifically it: 1) includes newly discovered radio
continuum sources, especially sources with a filamentary morphology
not identified in earlier, lower resolution, surveys; 2) covers
essentially the entire second quadrant along the Galactic plane; 3)
corrects misclassified and omitted objects in the \citet{fic86}
catalogue and 4) provides 408 MHz flux densities for the objects when possible.

\begin{figure*}
\centering
\begin{minipage}{140mm}
\includegraphics[width=140mm]{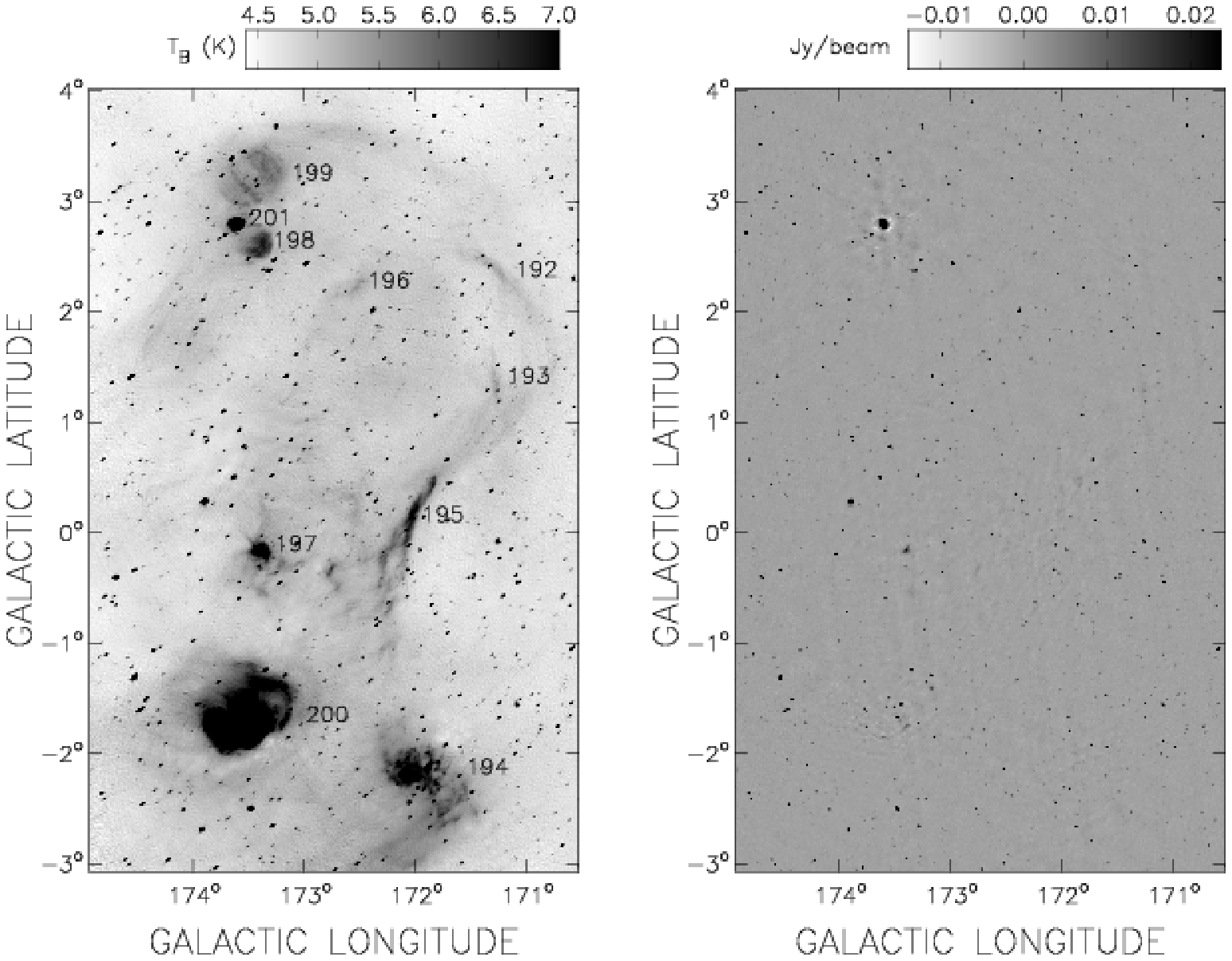}
\caption{The Galactic plane around $l=173\degr$ at 1420 MHz as imaged
  by the CGPS (left) and the NVSS all-sky survey (right). The CGPS observations
  recover the intense radio emission from the large number of
  \mbox{H\,{\sc ii}} regions in this region along with the extensive
  diffuse emission and a large number of filamentary structures. In
  contrast point sources dominate the NVSS image. Any extended
  emission has been resolved out leaving only a few small
  portions of the \mbox{H\,{\sc ii}} regions visible.}
\label{fig:nvsscomp}
\end{minipage}
\end{figure*}

In the next section the relevant characteristics of the CGPS data are
reviewed. Source identification and analysis techniques are discussed
in Section~\ref{sec:id}. The catalogue is presented and described in
Section~\ref{sec:cat}. Some interesting new radio sources and high
spatial dynamic range images of large star-forming complexes are
discussed in Section~\ref{sec:notes} followed by conclusions in 
Section~\ref{sec:conc}.

\section{Observations} \label{sec:observe}

As part of the CGPS, observations at 1420 and 408 MHz were
obtained using the Dominion Radio Astrophysical Observatory (DRAO)
interferometer \citep{lan00}. The first part of the survey (termed Phase 1)
covered approximately $75\degr < l < 145\degr$ between $-3\fdg5 < b <
+5\fdg6$ at 1420 MHz. The data acquisition and survey
strategy for Phase 1 are described in detail by \citet{tay03}. Phase 2 of
the survey extended the coverage along the Galactic plane to $62\degr < l <
175\degr$ using the same methodology as in Phase 1 of the survey.

The typical survey spatial resolution is approximately 1 arcmin at 1420 MHz
and approximately 3 arcmin at 408 MHz. Existing surveys from the Effelsberg and
Stockert single-dish telescopes were used to provide short-spacing
data \citep{rrf97, rei82, has82}. This, combined with the extensive
uv-plane coverage used in the individual interferometer observations,
means that the survey is sensitive to essentially all spatial
frequencies. This is crucial for the study of extended Galactic radio
emission as is illustrated by the comparison of the CGPS and NVSS (NRAO
VLA Sky Survey, \citealt{con98}) mosaics shown in Figure~\ref{fig:nvsscomp}.

\begin{figure*}
\centering
\begin{minipage}{140mm}
\includegraphics[width=140mm]{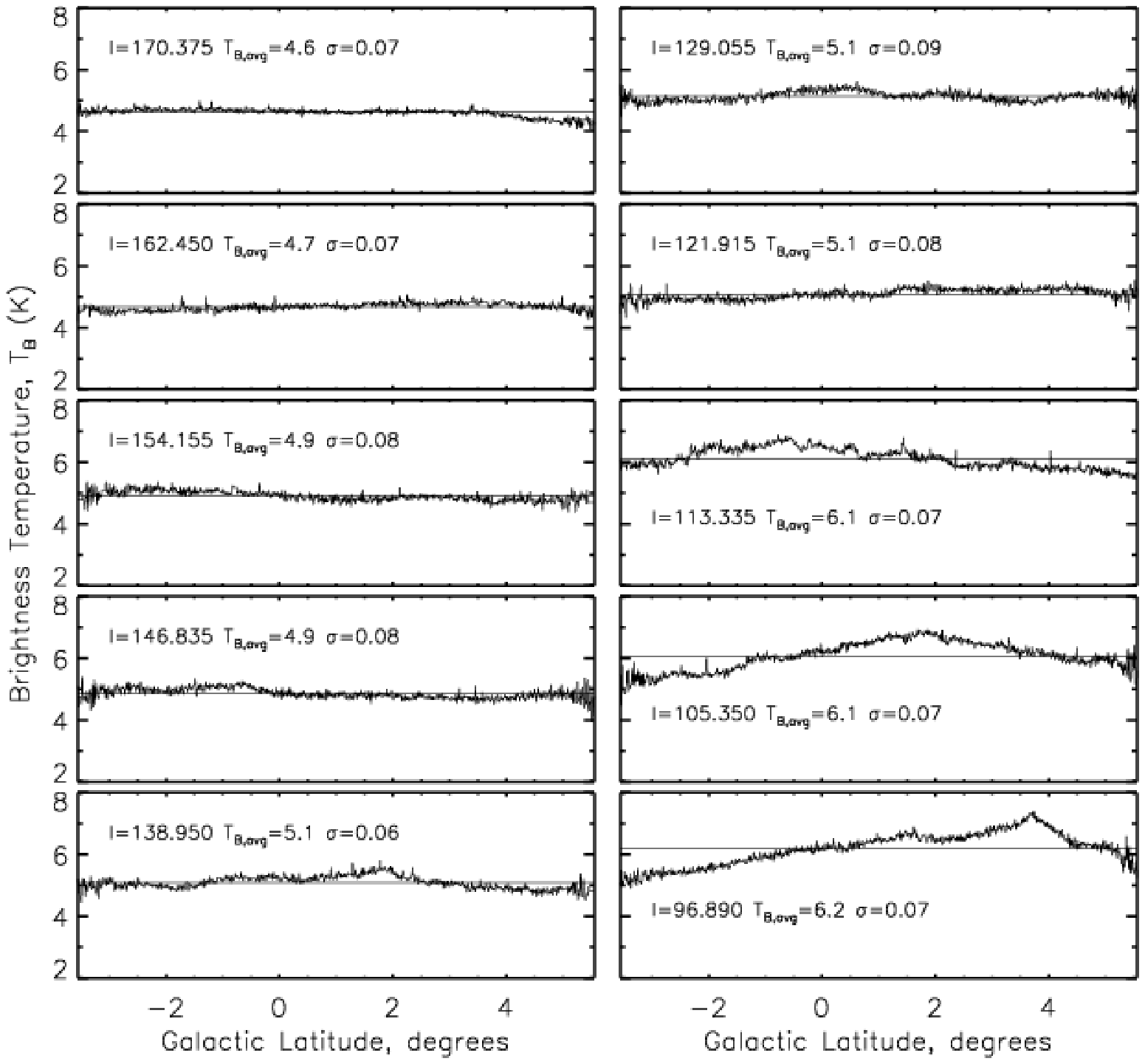}
\caption{CGPS Background Levels at 1420 MHz. Brightness temperature as
  a function of latitude is plotted at ten different longitudes
  spanning the survey region. The Galactic longitude, the average
  brightness temperature and the arcminute-scale noise level is shown
  for each profile.}
\label{fig:bg}
\end{minipage}
\end{figure*}

To illustrate the noise characteristics of the data at 1420
MHz we show in Figure~\ref{fig:bg} brightness temperature (T$_{\mathrm B}$) as
a function of galactic latitude at intervals spaced approximately 10
degrees apart across the survey region. The specific longitudes were
chosen to avoid extended sources and to minimize the number of point
sources in the cut. At the smallest ($\sim1$ arcminute) scale the measured
noise level is 1$\sigma = 0.07$~K which is in agreement with the
estimate quoted by \citet{tay03}. Also apparent in most of the cuts
is a very large scale (2 -- 5 degree) variation in the background with
a 0.25 -- 0.5~K amplitude. As noted by \citet{tay03} this large-scale
variation in the background limits our ability to identify faint
diffuse structures at these larger angular scales. 

CGPS data also includes \mbox{H\,{\sc i}} 21-cm line observations from DRAO,
$^{12}$CO ($J=1-0$) data from the Five College Radio Astronomy
Observatory (FCRAO) Outer Galaxy Survey (OGS; \citealt{hey98}) and
HIRES processed {\it IRAS} data from the {\it IRAS} Galaxy Atlas (IGA;
\citealt{cao97}) and the Mid-infrared Galaxy Atlas (MIGA;
\citealt{ker00}). All of these data sets have approximately 1 arcmin
resolution and, as part of the CGPS, are projected onto a common
spatial and (when appropriate) velocity grid (1024$\times$1024, 18
arcsec pixels, 0.82 km s$^{-1}$ channels). The resulting series of
$\sim 5\degr \times 5\degr$ mosaics or data cubes are freely available
via the Canadian Astronomy Data Centre (CADC).

\section{Source Identification and Analysis}\label{sec:id}

Each of the CGPS 1420 MHz $5\degr \times 5\degr$ mosaics for
$l>90\degr$ was visually inspected for extended radio sources using
the {\sc kvis} image viewer.\footnote{{\sc kvis} is part of the {\sc
karma} suite of image visualization tools \citep{goo96}.} The
inspection typically involved the use of a variety of greyscale
stretches in order to examine both faint and bright structures. Each
mosaic was viewed independently by at least two of the authors
before a final source list was compiled.

\begin{figure*}
\centering
\begin{minipage}{140mm}
\includegraphics[width=140mm]{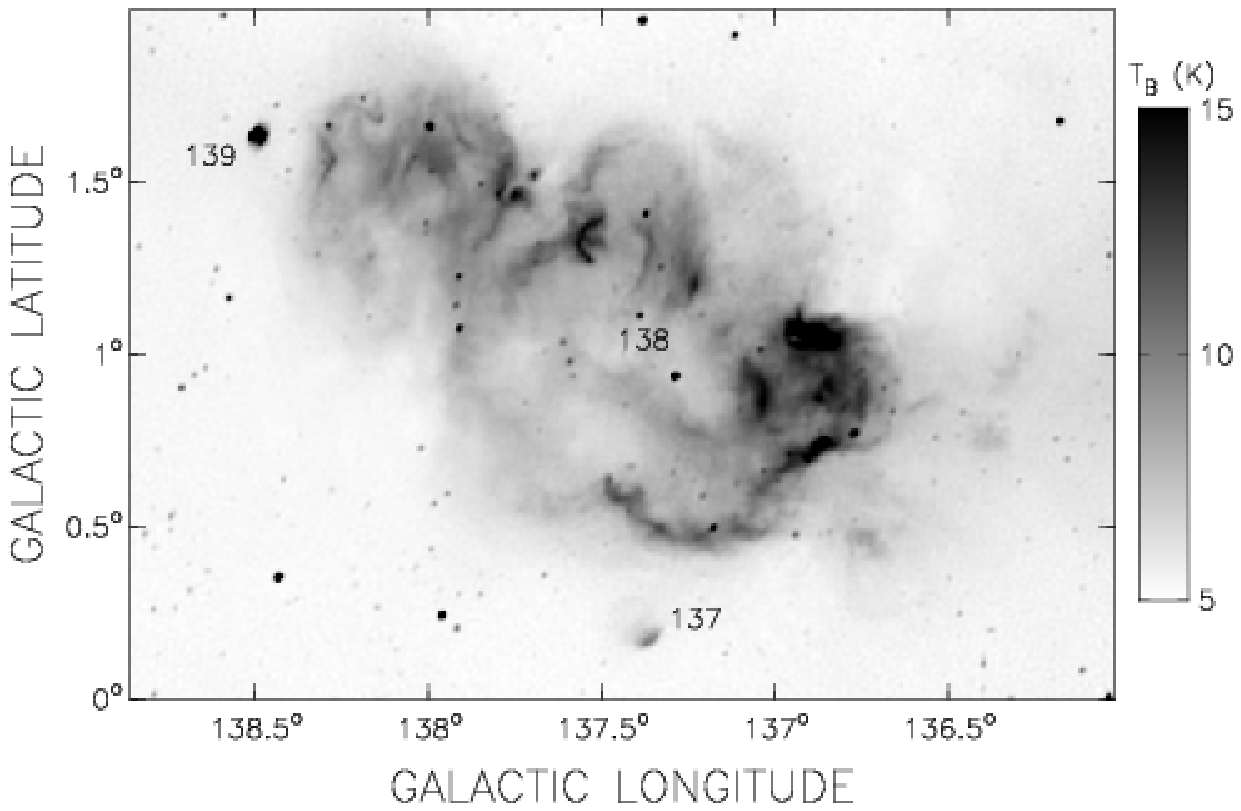}
\caption{CGPS 1420 MHz image of W5. This giant \mbox{H\,{\sc ii}}
  region is catalogued as a single extended source in this study along with a
  note indicating that the source has extensive
  substructure. Particularly striking in this image are the numerous
  bright rims, and the intense region of emission near $l=137\degr$,
  $b=+1\degr$. Also evident are regions of reduced intensity in the
  main body of the \mbox{H\,{\sc ii}} region probably associated with
  lower density regions cleared out by the stellar winds of the
  exciting O stars.}
\label{fig:w5}
\end{minipage}
\end{figure*}

We followed a grouping philosophy in the construction of the catalogue
where distinct substructures within extremely large sources were not
individually identified. For example, the numerous bright rims and
other distinct structures within the larger  W5 \mbox{H\,{\sc ii}}
region (see Figure~\ref{fig:w5}) were not individually noted, rather a
note was made in the catalogue that this source shows extensive
sub-structure. 

After sources were identified in the 1420 MHz mosaics
the corresponding 408 MHz mosaic was inspected to see if the source
could also be identified in the lower resolution 408 MHz images.
For each source the centroid position of the radio emission in
galactic coordinates was noted, the maximum and minimum axis of each
source (as seen at 1420 MHz) was measured using the {\sc kvis}
program, and a CGPS mosaic code was associated with each source.

\begin{figure}
\includegraphics[width=84mm]{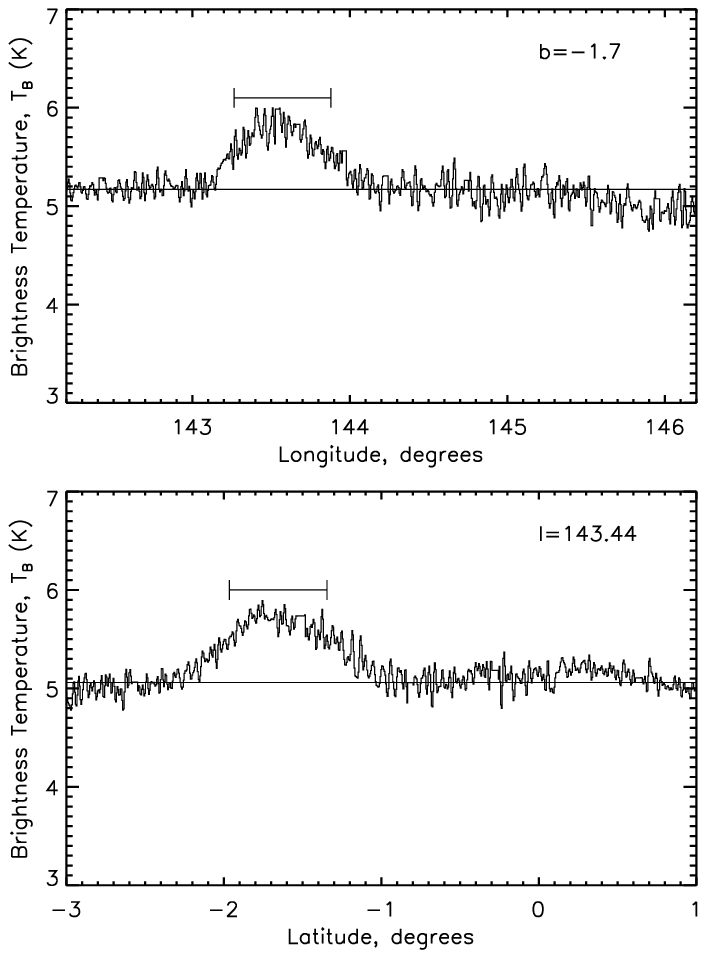}
\caption{CGPSE~153 T$_{\mathrm B}$ profiles. Profiles taken at
  constant latitude (top) and longitude (bottom) through the centroid
  position of CGPSE~153 are shown. The angular extent of the source, as
  determined by visual inspection, is indicated by the horizontal line
  above each profile, and can be seen to be $\sim0.4$~K above the
  local background (shown by the longer horizontal line in each
  panel). The noise level in both of the profiles is
  $\sim0.08$~K. Also apparent along these profiles is the low amplitude
  background variation referred to in Section~\ref{sec:observe}.}
\label{fig:153}
\end{figure}

Inspection of the extended sources identified in this manner showed 
that their extent was defined at a contrast level of $\Delta
$T$_{\mathrm B} = $ T$_\mathrm{edge} - $T$_\mathrm{bg} \simeq 0.3$~K, where
T$_\mathrm{edge}$ is the brightness temperature at the edge of the
region and T$_\mathrm{bg}$ is the local background level. This
contrast level is $\sim 4\sigma$ throughout the survey region.  
A typical example is shown in Figure~\ref{fig:153} where we plot
T$_{\mathrm B}$ at 1420 MHz along longitude and latitude cuts taken
through the centroid position of the large ($\sim 30$ arcmin scale),
faint (peak T$_{\mathrm B} \sim 6$~K) extended source CGPSE~153.

Flux density measurements were made at 1420 MHz and at
408 MHz (when possible),  using the {\sc imview}
program.\footnote{{\sc imview}  is part of the DRAO Export Software
Package}  This program  allows the user to interactively define a
background level using twisted-plane or twisted-quadratic fits to
user-selected background points. For the vast majority of the sources
the uncertainty associated with this background determination was the
largest source of error. The error estimates quoted in the catalogue
are the range of the flux density values determined using two
different, but equally valid, backgrounds. This situation is very
similar to that described by \citet{fic96} in their study of extended
infrared sources along the Galactic plane where background estimation
uncertainty also dominates the error budget of their flux density
measurements. The only exception to this was when the source was
isolated and particularly well defined. In these cases
uncertainties associated with original observations, 3\% and 6\% at
1420 and 408 MHz respectively \citep{kot06},are quoted. 

For SNRs located in the survey region the flux density measurements of
\citet{kot06}, which were obtained from the same data, are quoted. The
SNR catalogue of \citet{kot06} includes SNRs newly
discovered in the CGPS data and clarifies the nature of some of the
``possible SNRs'' listed in older versions of the \citet{gre06} SNR 
catalogue. In addition the very challenging flux density measurement
of the large diffuse \mbox{H\,{\sc ii}} region OA~184, again based on the same
data, was taken directly from \citet{fos06}.

\section{The Extended Source Catalogue} \label{sec:cat}

In this section we present the catalogue of extended radio sources in
the second quadrant of the Galaxy. The catalogue will be useful for
global studies of the distribution of SNRs and \mbox{H\,{\sc ii}}
regions within the outer Galaxy particularly along the Perseus and
Outer arms. An additional use of the catalogue will be in placing
higher-resolution studies in context. 

The catalogue is presented in three parts. The first part
(Table~\ref{tbl:cat1}) contains information on the position, angular
size and flux density of the sources. Column~1 is a running source
number (denoted CGPSE for CGPS Extended source). Galactic coordinates
of the source are given in Columns 2 and 3 followed by equatorial
coordinates (J2000) in Columns 4 and 5. The angular size of the source
is listed in Columns 6 and 7.  Columns 8 and 9 give the 1420 MHz flux
density and error estimate for each source. When possible the 408
MHz flux density and error estimates are given in Columns 10 and
11. For 58 of the sources 408 MHz emission could not be distinguished
from the background due to the higher 408 MHz noise level ($ 1\sigma
\sim 0.8$K) and lower resolution. This was particularly a problem in
the region around the Cas A SNR (the ME and MF CGPS mosaics) where
sidelobe effects can be observed at 408 MHz out to almost 5 degrees
from the SNR. In these cases no data values are indicated by
``$\cdots$'' in the printed tables and by flux density and error
values of ``999'' and ``99'' respectively in the machine-readable
version of the catalogue.  Finally, Column 12 gives the code of the
CGPS mosaic containing the source. Some sources are visible in
multiple mosaics because of the overlap between mosaics. In such cases
we selected the mosaic where the source is most centrally
located. Multiple codes in Column 12 are associated with those very
large sources that fall across mosaic boundaries.

The second part of the catalogue (Table~\ref{tbl:cat2}) provides
cross-identifications between the CGPSE sources and selected
catalogues. Associations were done by visually comparing, using {\sc
kvis}, the CGPS 1420 MHz mosaics with coordinates and size information
given for sources in various other catalogues. Column 1 lists the CGPSE
number. Columns 2 through 7 provide the applicable entry from the KR
catalogue, the SNR catalogue of \citet{kot06}, the Sh-2 optical
\mbox{H\,{\sc ii}} region catalogue \citep{sha59}, the RRF and F3R 
\citep{fur90} small-diameter radio source catalogues, and the
\citet{pal03} catalogue of Galactic \mbox{H\,{\sc ii}} regions
respectively. The final column lists other common associations along
with brief notes about the source in most cases. 

Finally, the third part of the catalogue (Table~\ref{tbl:cat3}) lists
the average local background and noise level for each of the sources. 
Column 1 lists the CGPSE number, Columns 2 and 3 are the background
and noise levels respectively at 1420 MHz and, when applicable,
Columns 4 and 5 list the same information for the 408 MHz data.

In total we identify 201 CGPSE sources, 27 of which have no
cross-references in the catalogues examined. The CGPSE catalogue is a
major update of the \citet{fic86} catalogue of extended sources as it
does not include KR sources that were incorrectly originally
identified as being extended (see \citealt{ker06} for more details). Also the
CGPSE catalogue includes some large objects that were not included in
\citet{fic86} (e.g., DA 568 as discussed in \citealt{ker04}) and
updates the nature of some of the mis-identified SNRs.

\begin{table}
\caption{CGPS extended sources -- Background Statistics}
\label{tbl:cat3}
\begin{tabular}{lcccc}
\hline
      & \multicolumn{2}{c}{1420 MHz} & \multicolumn{2}{c}{408 MHz} \\
CGPSE & $\mathrm{\overline{T}_{bg}}$ (K) & $1\sigma$ (K) &
      $\mathrm{\overline{T}_{bg}}$ (K) & $1\sigma$ (K)  \\
\hline
   1 &  7.70 &  0.10 & 72.89 &  0.89 \\
   2 &  7.85 &  0.10 & 71.08 &  1.10 \\
   3 &  5.03 &  0.14 & 42.82 &  1.14 \\
   4 &  8.05 &  0.17 & 78.71 &  2.58 \\
   5 &  8.50 &  0.21 & 80.51 &  3.91 \\
   6 &  8.09 &  0.21 & 77.33 &  2.25 \\
   7 &  6.93 &  0.07 & 68.65 &  1.64 \\
   8 &  6.73 &  0.11 & $\cdots$ & $\cdots$ \\
   9 &  8.01 &  0.09 & 79.75 &  1.17 \\
  10 &  7.59 &  0.13 & 74.83 &  1.08 \\
\hline  
\end{tabular}
\medskip

Table~\ref{tbl:cat3} is presented in its entirety in the electronic
edition of the journal.
\end{table}

Overlap between the CGPSE and the RRF and F3R catalogues is minimal
(there are $\sim 990$ RRF sources and $\sim 2000$ F3R sources within
the survey area) as the majority of the RRF and F3R sources remain
unresolved at the 1 arcmin resolution of the CGPS at 1420 MHz. Larger
CGPSE sources are also not found in the RRF and F3R catalogues as the
latter include only small diameter radio sources having angular extents of
$<16$ arcmin and $<12$ arcmin respectively. In some cases bright,
compact portions of larger CGPSE regions do have RRF or F3R identifications
and this is indicated in the notes to the source.

\section{High Spatial Dynamic Range Views of Star-Forming Regions} 
\label{sec:notes}

What truly sets the CGPS data apart from any previous observation of
our Galaxy at radio continuum wavelengths is its spatial frequency
sensitivity (see \S~\ref{sec:observe} and Fig.~\ref{fig:nvsscomp})
combined with its high spatial dynamic range, i.e., the combination of
hundreds of square degrees of spatial coverage with arcminute-scale
resolution. The unbiased nature of the survey also makes it well suited
for the discovery of new radio sources. In this section we highlight
three regions within the CGPS which illustrate the utility of the CGPS
data for large-scale synoptic studies of Galactic structure. In
addition to these examples we note a series of papers that have
already been published studying the region around $l=93\degr$
\citep{fos05,fos04,fos01}. This region also includes the enormous KR~1
(CTB~102) \mbox{H\,{\sc ii}} region which was highlighted in
\citet{ker06} and was recently the target of a radio-recombination line (RRL)
observing campaign at the Green Bank Telescope which will determine if
the extensive filamentary structure surrounding KR~1 is associated
with the main body of the \mbox{H\,{\sc ii}} region (Foster, Kerton \&
Arvidsson, in preparation).

\subsection{A new Galactic SNR candidate near $l=151\degr$,
  $b=+3\degr$}

\begin{figure*}
\centering
\begin{minipage}{140mm}
\includegraphics[width=140mm]{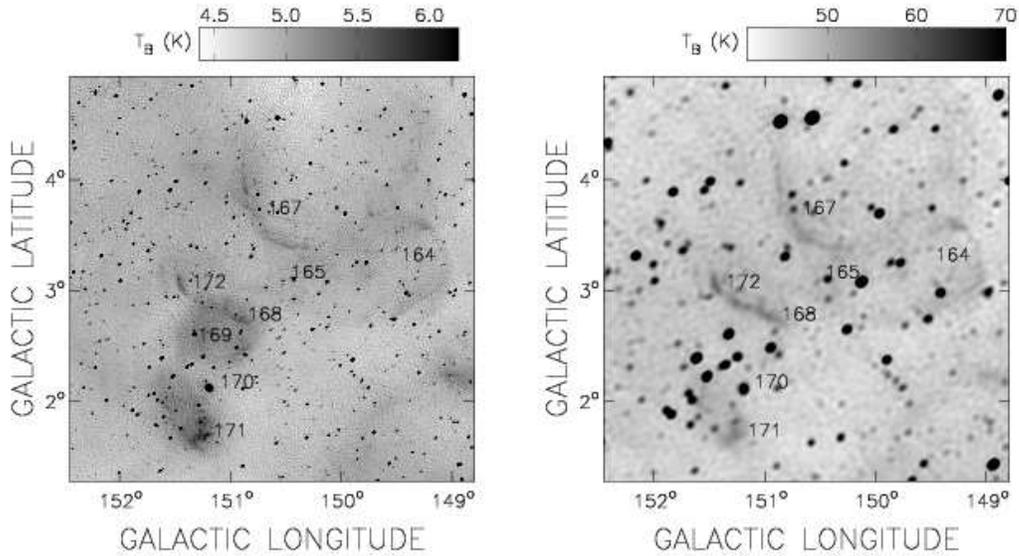}
\caption{A new Galactic SNR at 1420 (left) and 408 MHz (right). Extended radio
  sources are indicated by their CGPSE catalogue number. The filament
  traced by CGPSE 168 and 172 has a steep non-thermal radio spectrum
  making this a likely new Galactic SNR (G151.2+2.85)}
\label{fig:must}
\end{minipage}
\end{figure*}

Figure~\ref{fig:must} shows a 3.5 degree square region centered on
$l=150\fdg5$, $b=+3\degr$ at 1420 and 408 MHz. Except for the
smaller-diameter CGPSE~170 (Sh 2-207) and CGPSE~165 (F3R~4369) none of
the other extended sources in this region have been previously
noted. Table~\ref{tbl:mustalpha} summarizes the radio spectrum data
for the various extended sources in this region listed in Column~1 by
their CGPSE number.  Columns 2 and 3 give the spectral index between
408 MHz and 1420 MHz and the error estimate on the spectral index,
$\alpha_{408}^{1420}$ and $\sigma(\alpha_{408}^{1420})$
respectively. Here we use the convention that flux density ($F_\nu$)
and frequency ($\nu$) are related by $F_\nu \propto \nu^\alpha$. This is followed in Column~4 by brief notes on some of the sources.

Most of the extended sources have flat or rising spectral indexes ($|\alpha_{40
8}^{1420}| \leq 0.2$ or $\alpha_{408}^{1420} > 0.2$ respectively)
consistent with thermal radio emission associated with Galactic
\mbox{H\,{\sc ii}} regions. The filament CGPSE~167 also has associated
diffuse infrared emission, visible in the ancillary IR data sets
available as part of the CGPS, which would also be expected for a
Galactic \mbox{H\,{\sc ii}} region.

\begin{table}
\caption{Radio spectrum information for sources in the $l=151\degr$ region}
\label{tbl:mustalpha}
\begin{tabular}{lccl}
\hline
CGPSE & $\alpha_{408}^{1420}$ & $\sigma(\alpha_{408}^{1420})$  &  Notes \\
\hline
164 & $-0.2$   & $0.3$    &                            \\
165 & $-0.5$   & $0.05$   & extragalactic              \\
167 & $+0.06$  & $0.4$    & associated IR emission     \\
168 & $-1.3$   & $0.3$    & with 172, SNR     \\
169 & $\cdots$ & $\cdots$ & no 408 MHz detection    \\
170 & $-0.2$   & $0.2$    & optical \mbox{H\,{\sc ii}} region \\
171 & $+0.3$   & $0.2$    &            \\
172 & $-0.4$   & $0.2$    & with 168, SNR  \\
\hline
\end{tabular}
\end{table}

Three of the sources have falling (non-thermal, $\alpha_{408}^{1420} <
0.2$) spectra. CGPSE~165 is a likely extragalactic source.  It is only
slightly extended at 1420 MHz, has a non-thermal spectrum and has no
associated infrared emission. Particularly noteworthy though is the
filamentary structure made up of CGPSE 172 and 168. Both sections of
this filament have falling spectrum between 408 and 1420 MHz. A
combined flux density measurement yields a spectral index of
$-0.75$. This steep, non-thermal, spectral index combined with the
filamentary morphology seen at both 408 and 1420 MHz makes this object
a strong candidate for a new Galactic SNR (G151.2+2.85).

\subsection{A newly forming OB association at $l=112\degr$}

\begin{figure*}
\centering
\begin{minipage}{140mm}
\includegraphics[width=140mm]{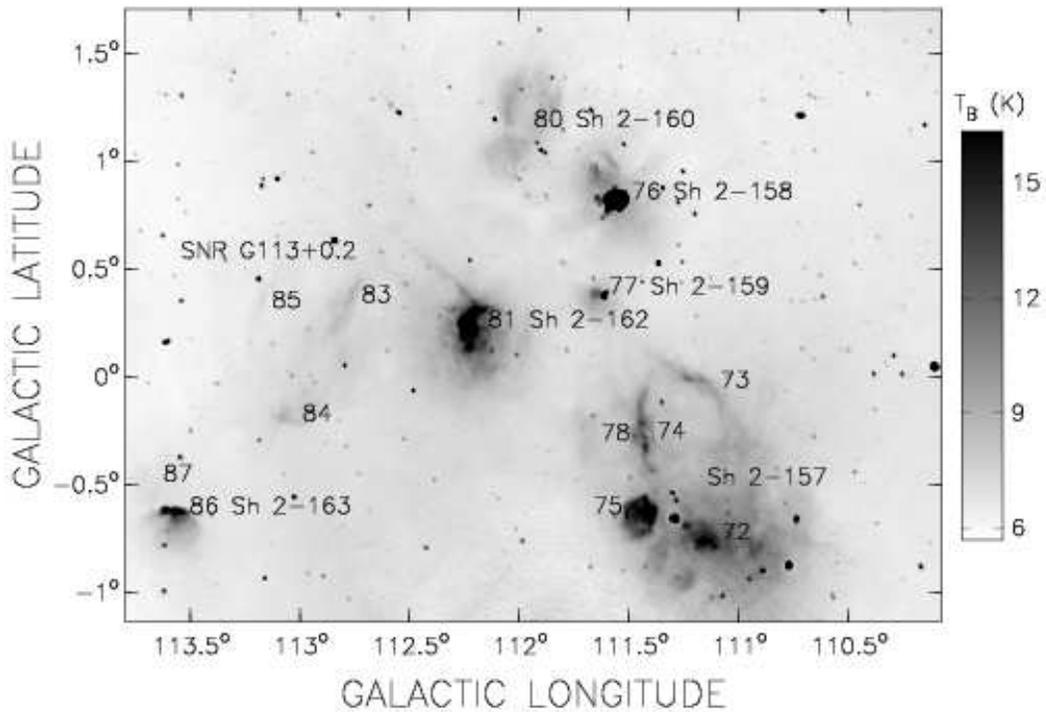}
\caption{The $l=112\degr$ complex. Artifacts from the Cas A SNR
  ($l=111\fdg735, b=-2\fdg130$) are faintly visible in the extreme
  lower portions of the image. Most of the emission is associated with
  optical \mbox{H\,{\sc ii}} regions (Sh 2 catalogue numbers are
  indicated along with the CGPSE numbers). CGPSE 73, 74, and 78 are
  also associated with Sh 2-157. The newly discovered SNR
  G113.0+0.2 is apparent as CGPSE 84, 85 and 86. CGPSE 87 is a faint newly
  identified extended radio source.}
\label{fig:l112}
\end{minipage}
\end{figure*}

Figure~\ref{fig:l112} shows the region near the Cas~A SNR. The region is rich 
in optical  \mbox{H\,{\sc ii}} regions and has been interpreted 
as being the early stages of the formation of an OB association
\citep{isr73,isr77} or even being part of a much larger complex of
star-forming regions \citep{loz86}. Some of the individual
\mbox{H\,{\sc ii}} regions have been observed at radio wavelengths at
higher resolution but the CGPS data show all of the regions in context
and are sensitive to the extensive extended emission. For example, the
observations of Sh-2~157 by \citet{isr77} and \citet{bir78} focus only
on the region around CGPSE 72 and 75, while CGPSE 73, 74, and 78 are
probably also related. The VLA survey of optically visible
\mbox{H\,{\sc ii}} regions by \citet{fic93} does not include any of
these regions presumably due to a combination of the angular size
limit for survey targets ($< 10$ arcmin) and the difficulty of
obtaining images close to the Cas~A SNR. 

While Cas A itself was not imaged by the DRAO interferometer because its
high brightness makes image calibration extremely difficult
\citep{tay03}, the CGPS image quality at 1420 MHz for the region close
to Cas A is extremely high. Special data reduction techniques,
described by \citet{wil99}, were used to minimize artifacts associated
with the bright SNR. In addition the field centres for the fields
surrounding Cas A were selected so that the SNR was located in the
first null of the primary beam of the DRAO interferometer at 1420 MHz 
\citep{tay03}. 

Since the CGPS observations were not directed only to the locations
of the optical \mbox{H\,{\sc ii}} regions a new Galactic SNR,
G113.0+0.2 (CGPSE~83, 84 and 85), was discovered at the high-longitude
end of this complex \citep{kot05}. Also CGPSE 87 is a newly identified
faint extended radio source. The combined infrared,
\mbox{H\,{\sc i}} line and high-resolution (approximately 45 arcsec)
molecular line data also available for this area would make the
reinvestigation of this region, in the context of it possibly being a
forming OB association, particularly worthwhile.

\subsection{The $l=173\degr$ complex}

There are a large number of optical \mbox{H\,{\sc ii}} regions with
sizes ranging from 0.2 degrees up to about 0.5 degrees located
around $l=173\degr$ (see Figure~\ref{fig:nvsscomp}). The emission
associated with these regions is encompassed by a very extended ``bowtie''
structure of which the southern portion is associated with
the large optical \mbox{H\,{\sc ii}} region Sh 2-230.

The region around CGPSE~201 (Sh 2-235) has been studied in some detail
\citep{isr78} owing to its compact nature, which facilitates
interferometric observations, and the fact it contains a number of
embedded infrared sources.  The region around the bright source
CGPSE~200 (Sh 2-236) is probably worthy of a more detailed investigation
beyond the existing studies which use this large \mbox{H\,{\sc ii}} as
as tracer of the Galactic velocity field (e.g., \citealt{bra93}).

The filaments CGPSE~192, 193, 195 and 196 are all apparently thermal
emission. CGPSE~195 has a rising spectrum between 408 and 1420 MHz and
infrared ({\it IRAS\/}) images of this area clearly show associated infrared
emission. CGPSE~196 is associated with a small ridge of infrared
emission which is originating from an embedded infrared source,
{\it IRAS}~05329+3628 within a distant molecular cloud
\citep{wb89}. There is no obvious infrared emission associated with
CGPSE 192 and 193, however, given the fact the filaments are not
detected at 408 MHz, they are most likely thermal emission associated 
with Sh 2-230.

\section{Conclusions} \label{sec:conc}

The CGPSE catalogue provides a summary of the extended radio emission sources
detected in the second quadrant of the Galaxy by the CGPS at 1420 MHz
and at  arcminute-scale resolution. In addition to flux density measurements
at both 1420 and 408 MHz, the catalogue contains positional
information and cross-identifications with other major radio and
optical \mbox{H\,{\sc ii}} catalogues. Also included are mosaic codes
for each source that will facilitate the retrieval of CGPS data from
the CADC. The catalogue includes a number of newly identified radio
continuum sources including a new Galactic SNR G151.2+2.85. It expands
the coverage of previous catalogues of extended Galactic radio sources
to include essentially the entire second quadrant along the Galactic
plane, and corrects a number of misclassifications and inadvertent
omissions of large sources from previous catalogues. For example,
CGPSE~34 and 37 (DA 558 and DA 568) were not included in the
\citet{fic86} catalogue, and objects like CGPS~17 and 187 are now
correctly classified as \mbox{H\,{\sc ii}} regions not SNRs.

The outer Galaxy is an ideal region for the study of the structure
of spiral arms and for the observation of the content of star-forming
regions owing to the relative proximity of the sources of interest and
limited confusion when compared with similar studies directed towards
the inner Galaxy. We have illustrated how the radio continuum data alone from
the CGPS can provide new synoptic views of star forming complexes and lead to
the discovery of new sources. The entire CGPS data set is a unique resource
for the investigation of Galactic structure and star formation and it is
hoped that this catalogue will facilitate the use of the CGPS data set
by the wider astronomy community.

\section*{Acknowledgments}
J.P. contributed to this research as part of the Freshman
Honors Program at Iowa State University. J.M. participated in this
research through an undergraduate research opportunity funded by Iowa
State University. The Dominion Radio Astrophysical Observatory is
operated by the National Research Council of Canada. The Canadian
Galactic Plane Survey is supported by a grant from Natural Science and 
Engineering Research Council of Canada.

\end{document}